\documentclass[12pt, fleqn]{article}
\usepackage[round]{natbib}
\usepackage{amsmath}
\usepackage{amssymb}
\usepackage{subcaption}
\usepackage{caption}
\usepackage{algorithm2e}
\usepackage{dsfont}
\usepackage{float}
\RestyleAlgo{ruled}
\usepackage{graphicx} 
\usepackage[nodisplayskipstretch]{setspace}
\usepackage[bookmarks=true]{hyperref}
\usepackage{comment}

\title{An early warning system for emerging markets\thanks{An early version of the paper was presented at iCEBA23, META24, EcoSta24 and ISF24, and the authors are grateful for feedback received from the participants. Research for this paper was supported by the Basic Research Program of HSE University (HSE-BR-2026-003).} 
}

\author{Artem Kraevskiy\thanks{National Research University Higher School of Economics, Moscow, Russia} \and Artem Prokhorov\thanks{The University of Sydney Business School, Sydney, Australia} \thanks{CEBDA, HSE University, Russia} \thanks{CIREQ, University of Montreal, Canada} \and Evgeniy Sokolovskiy\thanks{National Research University Higher School of Economics, Moscow, Russia} }
\date{April 2026}

\begin{document}

\maketitle
\sloppy

\begin{abstract}

\noindent Financial markets of emerging economies are vulnerable to extreme and cascading information spillovers, surges, sudden stops and reversals. With this in mind, we develop a new online early warning system (EWS) to detect what is referred to  as `concept drift' in machine learning, as a `regime shift' in economics  and  as a `change-point' in statistics. The system explores nonlinearities in financial information flows  and remains robust to heavy tails and dependence of extremes.  
The key component is the use of conditional entropy, which captures shifts in various channels of information transmission, not only in conditional mean or variance. We design a baseline method, and adapt it to a modern high-dimensional setting through the use of random forests and copulas. We show the relevance of each system component to the analysis of emerging markets. The new approach detects significant shifts where conventional methods fail. We explore when this happens using simulations and we provide two illustrations when the methods generate meaningful warnings. The ability to detect changes early helps improve  resilience in emerging markets against shocks  and provides new economic and financial insights into their operation. 

\end{abstract}
\setstretch{2}
\newpage

\section{Introduction}

Financial markets in emerging economies exhibit  unusually high volatility, frequent structural breaks, nonlinear dependence, asymmetry, and strong correlations in extremes, also known as tail-dependence 
\cite[see, e.g.,][]{CHAUDHURI:03, Chen/ibra:19, MAMONOV/etal:24}. Moreover, modern financial markets generally have a complicated structure of information flows often represented by  various global and local spillover and contagion effects, particularly strongly affecting emerging markets and manifesting themselves with significant lags and frictions in information transfer \cite[see, e.g.,][]{ZHAO/etal:22, KHALFAOUI/etal:23, AHMAD/etal:18}. In this setting, the rapid and reliable detection of shifts in the information channels at work in emerging markets has been identified as crucial for ensuring market stability, effective economic policy, and prevention of financial contagion \cite[see, e.g.,][]{NEAIME:16, SMIMOU:15, YANG:23}.  


In statistics literature, the area that develops principled ways of constructing early warning systems (EWS) which react to changes in a stochastic process with the shortest delay is known as change-point detection. This literature dates back to \cite{shiryaev:61, Shiryaev} and \cite{Roberts}, among others, and has accumulated an impressive set of optimality results \cite[see, e.g.,][]{Chen:19, Chu/chen:19, Pergamenchtchikov/tartakovsky:18, Pergamenchtchikov/tartakovsky:19}. Importantly, various versions of the so-called Shiryaev-Roberts (SR) statistic have been shown to be optimal for minimizing detection delay as the probability of a false alarm approaches zero. This applies to a wide range of stochastic processes, including autocorrelated time series of financial returns with unknown pre- and post-change distributions. This has spurred the development of online financial surveillance systems based on the SR statistic where ``online'' means real-time \cite[see, e.g.,][]{pepelyshev/poluchenko}.

Adapting these methodologies for use in emerging markets has been lagging behind. In econometrics, most of what is known as structural breaks methodology has been ``offline'', that is, it uses historical time series to find regime shifts \cite[see, e.g.,][]{HAMILTON:16, Bai-Perron, bai/perron:03}. The focus has been on obtaining consistent tests for the number and location of regime shifts in the available data, rather than on sequential monitoring for change-points. The conventional econometric methodology often assumes the number of regime shifts to be known or to be within a fixed region. It also imposes restrictions on the minimum duration of regimes and on distributional characteristics of time series such as finite higher-order moments. Such restrictions are particularly problematic when applied to volatile indicators of emerging markets.  

Moreover, traditionally macroeconomic analysis has focused on linear models of information transfer, such as \cite{granger:69} causality, vector autoregression (VAR), and dynamic conditional covariance (DCC) models. These models rely on parametric specifications and may suffer from significant  misspecification biases. Combined with the offline nature of such methods, their applicability in the setting of emerging economies seems questionable. At the same time, comparisons of performance between online and offline methods requires re-designing the offline methods to make them applicable in an online environment for which they were not intended. Therefore, there is little guidance in the literature on choosing between these methodologies.

Meanwhile, the field of machine learning has seen remarkable advances in areas related to change-point detection. Known as concept drift or data shift detection, this field of machine learning recognizes that the entire conditional distribution of the target variable given explanatory variables may change, drastically affecting predictive capacity of a machine learning engine trained on data from that distribution \cite[see, e.g.,][]{Hoens, Han}. This brings to the fore the recent successes of such advanced machine learning methods as random forests and artificial neural networks in estimating flexible non-parametric  conditional distributions and conditional expectation functions with many variable in the  conditioning set \cite[see, e.g.,][]{Breiman, Pospisil, Friedberg}.




Much relevant work is now happening in the intersection of machine learning and statistics. 
For example, \cite{Barros} develop a statistical estimator for concept drift detection based on Wilcoxon rank sum test. 
\cite{Baidari} 
apply  the  \cite{Bhattacharyya}  test to online concept drift detection. 
\cite{Wang} uses shifts in estimated covariance matrices to detect change-points, focusing attention on linear dependence; \cite{Ditzler} use shifts in the target variable distribution over time. \cite{Bifet} do the same but focus on developing a method 
of adaptive windowing (ADWIN) for high-frequency data stream monitoring. \cite{Raab} extend ADWIN to use the Kolmogorov-Smirnov statistic (KSWIN) and provide a more stable and precise estimation method. 
\cite{Tajeuna}  develop methods for finding regime shifts in multivariate time series via analyzing the co-evolving time series representation called mapping grid, while 
\cite{Yuxuan} and \cite{Alanqary} 
develop new ways of measuring  changes in statistical properties of multivariate distributions. \cite{JAMES/etal:23} develop a distribution-free market surveillance methodology based on dynamic time warping.  In spite of these impressive developments, none of the methodologies have been adapted to the settings commonly encountered in empirical analysis of emerging markets. 

The goal of this paper is to advance the early warning methodology by combining advances in statistics and machine learning with application in economics and finance, in particular in settings with highly volatile and tail-dependent data. Specifically, we design a new version of the change-point detection methodology based on conditional entropy. The novelty here is that information transfer is no longer assumed to be linear and it incorporates regime shifts in any feature of the distribution, not only in the conditional mean or variance. In fact, the mean and variance of the underlying series may not even be well-defined, which is not an uncommon setting in the analysis of extremely heavy-tailed financial time series, e.g., during financial crises and contagion periods. The EWS we design is sensitive to any shift in the conditional entropy estimated using state-of-the-art methods from machine learning. Conditional entropy relates the present state of the target variable to the information in the explanatory variables not captured by the lags of the target variable.  Therefore,  we directly measure the amount of uncertainty that cannot be explained either by the lagged target variable or by the explanatory variables, without restricting the relationships between the variable to be linear and parametric, and without focusing on a specific feature of the target distribution. 

As a starting point we take the conventional linear specification of the kind commonly used in empirical economics for Granger causality testing, and we convert it into a sequence of conditional entropy estimates. 
This permits an online sequential SR test based on entropy. 
To our knowledge, the use of conditional entropy in the SR statistic is novel and of interest  in its own right. We show how this approach helps model the stylized facts characterizing emerging markets data such as intensive information transmission, transfer termination and inversion \cite[see, e.g.,][]{Sahiner:24, Hassan:24}. 

Then, we develop extensions of this baseline setting that take advantage of local linear forests and of rank-based estimation. These extensions acknowledge the presence of  nonlinearities and tail-dependence in emerging markets data, and they permit infinite moments in the distribution of financial variables. By using ranks of the data, rather than the data itself, we introduce the concept of statistical copulas into the study of change-points, which, to our knowledge, has not been done before. Copulas capture rank-invariant nonlinear dependence in the data even if the marginal distributions are extremely heavy tailed and correlations and covariances are inappropriate dependence measures -- a feature that has prompted the growing popularity of copulas in modeling emerging markets data \cite[see, e.g.,][]{MAMONOV/etal:24, Tian:23, RODRIGUEZ2007401}. 

We present simulation results demonstrating that the new approach dominates a direct application of the SR test to the target variable or to AR process residuals in settings typical for emerging market studies. Specifically, we outline the cases when using ranks allows us to uncover regime shifts that cannot be detected by the conventional SR statistic and when the other methods of change-point detection fail or perform suboptimally. 
Additionally, the empirical section provides illustrations of how the proposed EWS could have issued alarms for the Russian equity market around the time of the Ukraine crisis and for the South African equity markets around the start of the COVID-19 recession.  

The paper is organized as follows. Section \ref{sec:prelims} sets the stage by defining notation and concepts such as entropy, random forests,  and change-points, highlighting their relevance to financial markets in emerging economies. Section 
\ref{sec:proc} introduces the baseline version of the proposed procedure and extends it using novel machine learning techniques like random forests and copulas to enhance change-point detection.  Section \ref{sec:sims} validates the proposed methods through simulations which  compare their performance against traditional techniques and alternative detection models. Section \ref{sec:appl} contains real data examples which illustrate how the EWS could have detected early warning signals in the Russian and South African financial markets around the time of major economic disruptions.  Section \ref{sec:concl} concludes. All codes and data are available at \url{https://github.com/kraevskiyAA/EWS_CondEnt}.

\section{Statistical preliminaries and related literature}\label{sec:prelims}

\subsection{Entropy and information transfer}

Suppose we have a target time series $Y_t$ and a history of explanatory variables available at time $t$, $\mathcal{X}_t = \{{X}_{t-1}, {X}_{t-2}, ..., {X}_{t-l}\}$, where each ${X}_{s}, s=t-1, \ldots, t-l,$ 
is a matrix containing $d$ explanatory variables (or their lags). As an example,  $Y_t$ can be a composite stock index of an emerging market and $\mathcal{X}_t$ can contain histories of commodity prices the country trades in. Our goal is to detect shifts in information transfer between the target and explanatory variables, causing an early warning about the state of the market described by the target variable.

It is common in information theory literature to evaluate the intensity of information transfer from $\mathcal{X}$ to $Y$ by \emph{conditional entropy of $Y$ given $\mathcal{X}$} \cite[see, e.g.,][]{ORLITSKY2003751} which can be written as follows:
\begin{equation}\label{eq:H(x|Y)}
    H = -\int\limits_{\mathbb{X},\mathcal{Y}} f(y, x) \cdot \log f(y|x) dydx = -\int\limits_{\mathbb{X}} f(x)\int\limits_{\mathcal{Y}} f(y| x) \cdot \log f(y|x) dydx = \mathbb{E}_{\mathbb{X}} \mathbb{E} \log f(y|x),
\end{equation}
where $f(y| x)$ is the conditional density of $Y$ given $\mathcal{X}$,  $f(y,  x)$ is the joint density of $(Y, \mathcal{X})$, and $f(x)$ is the marginal density of $\mathcal{X}$. The last equality follows from the definition of expectations and the law of iterated expectations.

Fundamentally, the concept of entropy measures the uncertainty about $Y_t$ given the information $\mathcal{X}_t$. In the context of signal transmission and unconditional distributions, \cite{hartley:28} is credited for observing that the information to be gained from $n$ outcomes each having a different probability $p_i$ can be defined as 
\[-\sum_{i=1}^n p_i \log_2 p_i.\]This has become known as the Shannon entropy \citep{shannon:48}. The higher this value, the higher is the uncertainty about the $n$ realizations.\footnote{A famous and straightforward example is a coin flip: the uncertainty is limited to two outcomes, head or tail. The information to be gained from a flip, known as a bit, is binary and can be defined by $\log_2 (2)=1$. Then, the information to be gained from $n$ flips is $\log_2(2)^n=n$. Intuitively, entropy is the level of uncertainty, or disorder, inherent in random outcomes.}  The largest amount of uncertainty corresponds to uniformly distributed outcomes, i.e., the same $p_i$ for all $i$. Eq.~(\ref{eq:H(x|Y)}) can be viewed as a generalization of this concept to continuous conditional distributions, showing the average amount of information to be supplied before one can communicate the information in $Y$ given that the other party knows the information in $\mathcal{X}$. \cite{Yang:18} provides an intuitive introduction to information theory for economists with a focus on entropy. 

\subsection{Entropy in emerging markets}

The conventional method of modeling and estimating the strength of information transfer in econometrics is via an ARMA, VAR or GARCH model, or their variations \cite[see, e.g.,][]{BEKAERT22}. These models focus on the first two statistical moments of the data, the mean and variance. There are several important problems with using these models for emerging markets. 

First, emerging markets data has much heavier tailed distributions than developed markets. For example,  \cite{ibragimov/etal:13} report that the tail index in the range 2.6-2.8 may serve as the boundary between the emerging and developed currency markets. They estimate that the moments of orders 2.6-2.8 and above are finite for the majority of developed countries, while they are infinite for most of the emerging countries. This suggests that a conventional application of central limit theorems in the estimation of volatility models for these markets is problematic because it requires finite fourth-order moments, which is impossible if moments of orders 2.6-2.8 are infinite.

Second, information transfer across markets often occurs through higher-order moments of the distribution or joint tail probabilities, rather than through the mean and variance. Changes in such information transfers cannot be adequately captured by standard methods, while such transfers are crucial to our understanding of how instability, contagion, and structural breaks are triggered and spread across emerging markets. For example, \cite{Yu/etal:20} document that the kurtosis coefficient of the Shanghai Composite Index (SHCI) between 1991 and 2016 is two orders of magnitude greater than that of the Dow Jones Industrial Average (DJI), and that there is evidence of tail-dependence between commodity and stock markets in China. Similarly, \cite{neslihanoglu:17} provide evidence of strong co-skewness and co-kurtosis patterns and structural breaks that distinguish large emerging stock markets.   

\begin{table}[h]
    \centering
    \caption{Beyond the first two moments: Entropy in emerging markets, 2021-2024}
    \begin{tabular}{lccccc}
        
        Market & Mean & Std. & Skewness & Kurtosis & Entropy \\
        \hline
        RTS Index & -0.001 & 0.081 & -0.008 & 12.213 & 2.211 \\
        UCI Index & 0.000 & 0.052 & -0.423 & 11.714 & 1.895 \\
        MERVAL Index & 0.005 & 0.031 & 0.156 & 4.906 & 1.827 \\
        BIST 100 Index & 0.003 & 0.026 & -0.851 & 7.793 & 1.510 \\
        \hline
        S\&P 500 Index & 0.000 & 0.012 & -0.256 & 2.669 & 0.927 \\
        EURO STOXX 50 & 0.000 & 0.012 & -0.523 & 4.619 & 0.882 \\
        \hline
    \end{tabular}
    \label{tab:entropy_markets}
\end{table}

The use of entropy bypasses these problems. Entropy is well defined regardless of the existence of moments. It captures variation in higher-order moments as well as in tail-dependence. Therefore, it is better suited to capture information transfer in emerging markets. To further illustrate this point, Table \ref{tab:entropy_markets} presents the mean, variance, skewness, kurtosis, and entropy estimates for logarithmic returns on the four stock market indices of emerging markets and two of developed markets. The data cover the period from the first trading day in 2021 to the first trading day in 2024, and only the trading days where all six indices were observed at the same time. The four emerging markets are 
Russia (RTS), Uzbekistan (UCI),  Argentina (MERVAL) and Turkey (BIST 100). 
The two developed markets are the United States (S\&P500) and Eurozone (Euro Stoxx 50), for which the indices track the stocks of 500 largest US companies and the stocks of 50 Eurozone companies, respectively.

It is clear from the table that, even during this relatively quiet period, the emerging markets exhibit a substantially higher degree of asymmetry, volatility, and kurtosis than the developed markets. 
The kurtosis coefficients are several times higher for emerging markets than for developed markets, for which they are not too far from 3 -- the value corresponding to the normal distribution. Importantly, we can see that the entropy captures the various distinguishing aspects of the return distribution in emerging markets. For example, it captures the excess kurtosis of the four emerging market indices. 

\subsection{Non-parametrics and random forests}

An important component of computing entropy $H$ is the estimation of conditional density. A flexible way to estimate $f(y|x)$ without making strong distributional assumptions is to use tools from non-parametric econometrics \cite[see, e.g.,][]{li/racine:23}. However, most standard methods will not work with conventional sample sizes when the dimension of $\mathcal{X}_{t}$ is even moderately large. This problem is known as the curse of dimensionality. Additionally, non-parametric methods are known to perform poorly in the extremes of the distribution, which are of particular interest when dealing with data from emerging markets.

An alternative approach that has been gaining popularity but has not been explored in the context of entropy-based EWS is to use machine learning techniques. As a prominent example, \cite{Pospisil} recently proposed using random forests for conditional density estimation (RFCDE). The estimator improves on the standard non-parametric method of kernel density estimation (KDE) by employing adaptive weights obtained using random forests of \cite{Breiman}. The main advantage here is that random forests break the curse of dimensionality by using random subsets of the entire set of conditioning variables. 
This allows RFCDE to achieve consistent conditional density estimation with low computational costs and high accuracy.

We use the RFCDE to obtain consistent estimates of $f(y|x)$. Then, it is intuitive that a spike in $H$ evaluated using the RFCFE of $f(y|x)$ should indicate an increase in uncertainty (a decrease in information transfer from $\mathcal{X}$ to $Y$) and the remaining question is how to identify significant spikes. We note that the RFCDE has not yet been used to estimate conditional entropy and offers a new way to evaluate information transmission in high-dimensional settings. 

\subsection{Dimensionality and misspecification in models of emerging markets}

When modeling the dynamics of emerging markets, the issues of dimensionality and model misspecification are particularly acute. This is due to a larger number of factors that can be overlooked while they have a stronger impact on emerging economies than on developed ones. Such factors include external financial indicators capturing spillover effects, geo-political events, economic and financial reforms including various capital control measures, macroeconomic stabilization programs, large-scale privatization, and so on. For example, \cite{CEPNI:19} use 88 to 117 economic time series to build predictive models for the Turkish, Brazilian, Mexican, South African, and Indonesian economies and report various significant spillover effects across four out of the five emerging market economies.

For the same reasons, it has been harder to argue for correct specification of dynamic models of emerging markets without having to increase the model dimensionality or resorting to non-parametrics. For example, \cite{BEKAERT22} augment their bivariate VAR of returns and equity flows for 20 emerging economies by including various additional effects of market capitalization, global interest rates, local dividend yields and structural reforms on equity flows. Incorrectly specified models lead to biased and inconsistent estimators,  which invalidate any potential policy recommendations based on the models' predictions.

\subsection{Change-point detection}

Traditional change-point detection literature, going back more than 60 years \citep{page:54, Shiryaev, Roberts}, looks at an aggregated sequence of likelihood ratios based on the likelihood of a change at time $t$ versus the likelihood of no change, over a history of observations, where the densities before and after the change are known but the change-point is not. A recent semiparametric application to spillovers in conditional correlations between developed and emerging markets is \cite{Barassi/etal:20}.  There now exist important relaxations of the assumptions of known densities and operational methods that look at a sliding window of data rather than the entire history. However, this literature has not looked at entropy- or rank-based evaluation of information transfer, not to mention the setting of emerging markets.      

Define the Shiryaev-Roberts statistic as follows:
\begin{equation}
    SR_t = \sum\limits_{k=1}^{t}\prod\limits_{i=k}^{t}\dfrac{f_{0}(\xi_i)}{f_{\infty}(\xi_i)},
\end{equation}
where $\xi_i, i=1,\ldots, t,$ are generic independent random variables or vectors, $f_0$ is the density of $\xi_i$ before the change and $f_\infty$ is the density of $\xi_i$ after the change. Then, it is easy to derive the recursive version
\begin{equation}
    SR_t = (1+SR_{t-1})\frac{f_0(\xi_t)}{f_{\infty}(\xi_t)}.
\end{equation} 

We use a version of  the weighted SR statistic proposed by \cite{Tartakovsky}. Their procedure permits randomness in the drift of $\xi_t$ in the sense that at each $t$ the change may take one of $m$ values and the weighted statistic has the form
\begin{equation}
        SR_t^{w} = \frac{1}{m}\sum\limits_{i=1}^{m}  SR_t(\Delta\mu_i),
\end{equation}
where  $SR_t(\Delta\mu_i)$ is the value of $SR_t$ when the difference between the mean of $\xi_t$ before and after the change is $\Delta\mu_i$.\footnote{We use $m=6$  shifts in the mean with a fixed step $\Delta \mu$ equal to the range of data divided by m.} This leads to a much better performance in cases when distribution  parameters are unknown. Besides that, the construction of the statistic $SR_t$ is standard \cite[see, e.g.,][]{pepelyshev/poluchenko}.

An alarm will sound when $SR_t$ exceeds a threshold value that controls the false alarm rate and, according to a series of important results in statistics, the SR procedure is optimal in the sense of minimizing the average detection delay \cite[see, e.g.,][]{poluchenko/tartakovsky:10}. 

\subsection{Change-points in emerging markets}

Early change-point detection is crucial for predicting and analyzing financial crises, contagion, speculative bubbles, and business cycles \cite[see, e.g.,][]{Greenwood/etal:22}. For example, \cite{WHITEHOUSE:25} recently used a new change-point detection techniques to show that the United States housing bubble -- the precursor for the global financial crisis -- could have been detected as early as the first quarter of 1999. This could have helped prevent the crisis from spreading and mitigate the fallout. 

With reference to emerging markets, change-points, also referred to as structural changes, remain arguably the most important sources of potential misspecification in dynamic econometric models. An incorrect account for change-points in an emerging market associated with structural reforms, geo-political and other factors, leads to nonstationarities that invalidate econometric analysis.  From the empirical perspective, early detection of change-points helps identify what has become known in the literature as \emph{terminations} (or \emph{sudden stops}) and \emph{reversals} of capital flows observed in emerging markets \cite[see, e.g.,][]{Calvo:98, SARMIENTO:24}.  

Given that much information transmission in emerging economies may be occurring in higher-order moments rather than in the mean and variance, and may involve tail-dependent multivariate distributions with heavy-tailed marginals, the conventional SR statistics based on the known densities $f_0(\xi_t)$ and $f_{\infty}(\xi_t)$ are likely to miss relevant change-points. 


\section{The new procedure}\label{sec:proc}

The core of the new procedure is an application of the Shiryaev-Roberts statistic to a sequence of entropy estimates of $H$ assessed on a moving window of length $\Delta$. 
The window length needs to be short enough so that the regime shift is noticeable but long enough not to pick noise (we use window lengths of 50 to 100 observations in practice).
An EWS signal is  triggered when $SR_t^w$ evaluated using the entropy process $H_t$ surpasses a threshold $A>0$, indicating a significant shift in information transmission from $\mathcal{X}_t$ to $Y_t$. 

\subsection{Linear projections and Granger causality}\label{sebsec:granger}

We start by fitting an autoregressive model for the target variable $Y_t$ on a moving window of length $\Delta$: 
    \begin{equation}
        \hat{Y_t} = \hat{\alpha}_0 + \sum\limits_{s=1}^{k} \hat{\alpha}_i Y_{t-s},
    \end{equation} 
where we use the Bayesian Information Criterion (BIC) to determine the optimal AR order $k$. This gives the part of $Y_t$ orthogonal to the history of the target variable. This is the unexpected component of $Y_t$ we are interested in, which we write as follows: 
    \begin{equation}\label{eq:resid}
        e_t = Y_t - \hat{Y}_t.
    \end{equation}
The AR(k) residuals $e_t$ capture any remaining news in the target variable $Y_t$ and we wish to explore how the distribution of this variable changes depending on the explanatory variables $\mathcal{X}$. We wish to remain agnostic about the specific moments of the distribution that change and about the form of the distribution. 

As the next step, we obtain the fitted value from the regresion of $e_t$ on lags of the explanatory variables contained in $\mathcal{X}$ using the same moving window: 
\begin{equation}\label{eq:granger}
        \hat{e}_t = \hat{\beta}_0 + \sum\limits_{j=1}^{d}\sum\limits_{s=1}^{l} \hat{\beta}^{(j)}_{s}X_{t-s}^{(j)},
    \end{equation}
where $X_{t-s}^{(j)}$ is one of the $d$ available explanatory variables, lagged $s$ times, $s=1, \ldots, l, j=1, \ldots, d$. Again, the optimal lag $l$ is selected by BIC. 

Clearly, these two steps are equivalent to a one-step regression of $Y_t$ on both its own history and history of the covariates. This corresponds to the familiar concept of (linear) Granger causality, which could be detected by a joint significance test of $\hat{\beta}^{(j)}_{s}$. 

A recent example of Granger causality analysis for emerging markets is \cite{balcilar/etal:22} who implement a mixed-frequency version of these two steps to establish if economic policy uncertainty Granger causes GDP growth of seven emerging market economies while controlling for the effect of oil price, interest rates, and CPI. They document strong evidence for (linear) Granger causality in Brazil, Chile, and India and weak evidence for Colombia, Mexico, and Russia. We note that these results are valid insofar as the data used for the test is not too heavy tailed. 

\subsection{Baseline version of entropy}\label{subsec:baseline}

In empirical finance, it has been noted at least since \cite{hiemstra/jones:94}, that linear causality is too restrictive. A standard response has been  to include nonlinear functions of the explanatory variables. However, this limits the permitted nonlinearities to those specific functional forms of $X$ that are included in the second step. 

In the baseline version of the new procedure, we go beyond the conventional notion of Granger causality by computing the conditional entropy of $e_t = Y_t -\hat{Y}_t$ given the second step residuals $\hat{e}_t$. This recognizes the fact that change-point detection should be based on the conditional distribution of the new information in $Y_t$ (not captured by its own history) given the history of $X_{t}^{(j)}$ (captured by $\hat{e}_t$). In other words, the density $f(e_t|\hat{e}_t)$ is more informative than the linear regression $\mathbb{E}( e_t|\hat{e}_t)$ since it accommodates more complex (than linear in mean) patterns of dependence between $Y$ and $X$. 

Let $\hat{f}$ 
denote the RFCDE of the conditional density of $e$ given $\hat{e}$ obtained using the moving window of size $\Delta$. Let  $\hat{\mathcal{E}}_{[t_0, t_1]} = \{\hat{e}_{t_0}, \hat{e}_{t_0+1}, ..., \hat{e}_{t_1}  \}$ denote the set of $\hat{e}$'s that fall into the window. Then, the estimator of aggregate information transfer over the window can be written as a sum of the entropy contributions from the values of $\hat{e}$ that fall into the window: 
 \begin{equation}\label{eq:simpson}
        \hat{H}_{[t_0, t_1]} = \sum\limits_{\hat{e} \in \hat{\mathcal{E}}_{[t_0, t_1]}} \bigg[\int\limits_{\mathbb{R}}  \hat{f}(\varepsilon| \hat{e}) \cdot \log \hat{f}(\varepsilon| \hat{e}) d \varepsilon\bigg].
    \end{equation}
We can approximate the integrals in (\ref{eq:simpson}) using stochastic or deterministic integral approximation, e.g., Simpson's rule, which is a deterministic method and which is our preference in simulations and applications. We are interested in early identification of spikes in $\hat{H}_{[t_0, t_1]}$ as we move across time. 

Denote the sequence of entropy estimates over a sliding window of size $\Delta$ as follows
    \begin{equation*}
        \hat{H}_{[t_0, t_0 + \Delta]}, \hat{H}_{[(t_0+1), (t_0+1) + \Delta]}, ...
    \end{equation*}
We note that the step between consecutive values of $H$ in the sequence can be larger than one time period but this will delay detection while expediting computation. 

\subsection{Entropy based algorithm}

The sequence $ \hat{H}_{[t_0, t_0 + \Delta]}, \hat{H}_{[(t_0+1), (t_0+1) + \Delta]}$ is the time series to which we apply the SR procedure in the baseline version of our approach. For concreteness, we provide the detailed pseudo-code in Appendix \ref{app:algo}. The algorithm uses several inputs besides the sequence of $H$, namely, the set of shift values $M_{\Delta}=\{\Delta\mu_1, \ldots, \Delta\mu_m\}$, the threshold $A$ and the smoothing parameters $\alpha$ and $\beta$.  As outputs, it produces change-points $\tau$. 

The performance of the proposed procedure depends on several tuning parameters. The window length $\Delta$ controls the trade-off between estimation precision and detection delay: shorter windows react faster to changes but produce noisier entropy estimates, while longer windows improve stability at the cost of slower detection. In our experiments, values in the range $\Delta \in [50,100]$ provide a reasonable balance.

Smoothing parameters $\alpha$ and $\beta$ are used at the end of each iteration of Algorithm \ref{alg:SR} in order to produce an exponential smoothing update of the conditional expectation and variance of $H$, balancing the past values and the new information. The natural value for $\alpha$ is 0.5, reflecting equal weights on the old value of the mean $\hat{\mu}_{t-1}$ and the current $H_t$, while for the variance update, we choose $\beta = 0.9$, putting more weight on the estimate of $\hat{\sigma}$ from the previous period and thus dampening the variance update. Larger values of $\alpha$ and $\beta$ result in smoother trajectories of the moments of $H$ but may delay detection.

The threshold $A$ determines the false alarm rate. Although there is no strict rule for selecting the value of threshold $A$, it is customary to consider two minimization criteria, namely, average detection delay (ADD) defined as $   \mathbb{E}_{\theta}[\tau-\theta|\tau>\theta]$ and probability of false alarm (PFA) defined as $ \mathbb{P}_{\theta}( \tau < \theta)$, where $\theta$ is the actual moment of the change and $\tau = \arg\min_{t} \{t|SR_t > A\}$  is the moment of time it is detected. 
\cite{Pergamenchtchikov/tartakovsky:18, Pergamenchtchikov/tartakovsky:19} show that the weighted SR procedure approximately minimizes ADD  among all change-point detection procedures with a given PFA within a window of a fixed length for non-i.i.d.~data, including autoregressions and conditional heteroskedasticity GARCH-type models. 

However, these quantities are generally unavailable to the researcher in practice, unless one deals with synthetic data and is able to generate variables using a given constellation of these parameters. For instance, in Section \ref{sec:sims} where we report simulation results, we are able to set $A$ by choosing the values of ADD and PFA via the numerous relationships derived in the literature \cite[see, e.g.,][]{Tartakovsky_th}. In Alternatively, $A$ can be selected using (i) simulation-based calibration to target a desired probability of false alarm, or (ii) a grid search combined with sensitivity analysis. In empirical applications, we recommend reporting results for a range of thresholds to assess robustness.

%




\subsection{Nonlinearities in financial market data}

The baseline entropy computation described in the previous subsection uses a conditional density estimator of the AR process residuals ${e}_t$ given $\hat{e}_t$, where $\hat{e}_t$ represents the linear span of the lagged explanatory variables $X_{t}^{(j)}$ in Eq.~(\ref{eq:granger}). This aligns with the traditional view of the Granger causality, which identifies information transfer by testing joint significance of the lagged regressors in a linear autoregressive model. In the previous sections we constructed  a flexible, kernel-based estimator $\hat{f}(e_t|\hat{e}_t)$ and applied the SR statistic to a moving window of entropy estimates $\hat{H}_{[t_0, t_1]}$ obtained using $\hat{f}(e_t|\hat{e}_t)$. However, this does not take full account of nonlinearity. 

Meanwhile, nonlinearities in financial market data have been widely documented. This has been the case for developed markets for a while and evidence has been accumulating recently for emerging markets, as well. For example, \cite{caselli/roitman:19} document nonlinearities and asymmetries in the transmission of exchange rate fluctuations to prices for 27 emerging markets while \cite{Hasanov/omay:08} found strong nonlinearities in stock returns in Greece and Turkey. Interestingly, \cite{Guhathakurta:16} apply a battery of nonlinearity tests to both developed and emerging stock markets in order to compare them in terms of prevalence of nonlinearities. They find that all the time series considered, whether from developed or emerging markets, exhibit significant nonlinearities. Needless to say, this has important repercussions for  many areas in empirical finance, e.g., for empirical tests of market efficiency, purchasing power parity, tests of stationarity, cointegration, causality and empirical models of asset pricing which have relied on linearity. 

To better account for nonlinearities, we extend the baseline entropy calculation in two directions: random forests and copulas. 


\subsection{Local linear forest}\label{subsec:llf}

The univariate conditioning used in the density estimate $\hat{f}(e | \hat{e})$ in Section \ref{subsec:baseline} substantially simplified the task of non-parametric density estimation. The alternative of conditioning on the entire set of covariates $X_{t-s}^{(j)}$, without obtaining $\hat{e}_t$, leads to a non-parametric estimator $\hat{f}(e_t | \mathcal{X}_t)$ which suffers from a severe curse of dimensionality due to the high dimensionality of $\mathcal{X}_t$, making the estimate very imprecise. The problem is further  exacerbated by the trade-off between quick detection delay and accuracy of density estimation. It is well known in the change-point detection literature that if we increase the rolling window, making the density estimation more precise,  the time before change-point detection increases. Thus, the standard multivariate non-parametric detection procedures are either inaccurate or slow. 

To overcome this issue without increasing the dimensionality of density estimation, we use a recently developed method of local linear forest  (LLF) proposed by \cite{Friedberg}. LLF constructs an estimate of the conditional mean $\tilde{e}_t= \mathbb{E}[e_t|X_{t-1}=x_{t-1},\ldots, X_{t-l}=x_{t-l}]$ achieving a faster convergence rate than traditional non-parametric estimators of the same dimensionality, and preventing the overfitting. The standard (non-local) random forest estimator \cite[see, e.g.,][]{Breiman, Probst} is sensitive to hyperparameter  selection and tends to overfit. The use of LLF improves upon the linear Granger causality regression (\ref{eq:granger}) and is new to financial econometrics.

Specifically, our first extension of the baseline methodology is to use the flexible LLF estimator $\tilde{e}_t$ instead of the fitted value $\hat{e}_t$ in constructing $\hat{f}(e_t | \tilde{e}_t)$. The SR statistic is now based on a sequence of the new estimates of $H$ on a sliding window $[t_0, t_1]$:
\begin{equation}
    \hat{H}^{LLF}_{[t_0;t_1]} = \sum\limits_{\tilde{e} \in \Psi_{[t_0; t_1]}} \bigg[ \int\limits_{\mathbb{R}} \hat{f}(\varepsilon| \tilde{e}) \cdot \log{\hat{f}(\varepsilon| \tilde{e}) d\varepsilon}\bigg],
\end{equation}
where $\Psi_{[t_0; t_1]} = \{\tilde{e}_{t_0}; \tilde{e}_{t_0+1}; ...; \tilde{e}_{t_1}\}$. The rest of the methodology remains unchanged.

We note that both $\hat{H}_{[t_0, t_1]}$ and $\hat{H}^{LLF}_{[t_0, t_1]}$ permit nonlinearity in the relationship of $Y_t$ on $\mathcal{X}_t$ because both use RFCDE to construct the conditional density of $e_t$. The advantage of using the new estimates $\tilde{e}_t$ is that it implies a nonlinear conditional mean of $Y_t$ in terms of $\mathcal{X}_t$, given lags of $Y_t$. Therefore, it aligns with the notion of nonlinear Granger causality where $\mathcal{X}_t$ Granger-causes $Y_t$ if the non-parametric estimator  $\tilde{e}_t$ is indistinguishable from a constant in each dimension of $\mathcal{X}_t$. Unfortunately, powerful statistical tests of this null hypothesis in a setting with many covariates in $\mathcal{X}_t$  are not available. Importantly, the procedure we propose does not require such 
testing. 

\subsection{Copulas}

Our second extension addresses the problem of heavy tails and tail-dependence, in addition to the problem of nonlinearity. It is well documented that financial markets of emerging economies are characterized by asymmetric heavy-tailed distribution, which make application of standard econometric methods problematic \cite[see, e.g.,][]{Chen/ibra:19}. As an alternative we follow the growing literature on copula-based estimation in finance \cite[see, e.g.,][]{Ibragimov/prokhorov:17} and re-formulate the problem of density estimation using ranks instead of actual values.

In the proposed modification, the baseline RFCDE estimator $\hat{f}(e_t|\hat{e}_t)$ is replaced by the rank-based RFCDE estimator $\hat{f}(u_t|\hat{u}_t)$, where 
\begin{equation*}
    u_{t} = \dfrac{1}{\Delta} \sum\limits_{s=1}^{\Delta} \mathds{1}\{e_{t-s} \leq e_t \}, \qquad \hat{u}_{t} = \dfrac{1}{\Delta} \sum\limits_{s=1}^{\Delta} \mathds{1}\{\hat{e}_{t-s} \leq \hat{e}_t \},
\end{equation*}
and, as before, $\Delta$ is the length of the sliding window over which we estimate the densities. The values of $u_t$ and $\hat{u}_t$ are known as pseudo-observations; they are values between $[0,1]$ corresponding to the ranks of observations $e_t$ and $\hat{e}_t$, respectively, in the sliding window, divided by the number of observations in the window.
The corresponding entropy estimates can be written as follows
\begin{equation}\label{eq:H_ranks}
    \hat{H}_{[t_0;t_1]}^{rank} = \sum\limits_{{\hat{u}} \in \mathcal{U}_{[t_0; t_1]}} \bigg[\int\limits_{0}^{1} \hat{f}(w |{\hat{u}}) \cdot \log \hat{f}(w |{\hat{u}}) dw\bigg],
\end{equation}
where $\mathcal{U}_{[t_0; t_1]} = \{\hat{u}_{t_0}; {\hat{u}}_{t_0+1}; ...; {\hat{u}}_{t_1} \}$ and  $\hat{f}(u |{\hat{u}})$ can be interpreted as the conditional copula density of $e_t$ given $\hat{e}_t$.
The remaining parts of the procedure are unchanged.\footnote{We also used the LLF estimates $\tilde{e}_t$ instead of $\hat{e}_t$ but do not report this case for brevity.}

\subsection{Relevance of rank-based models to emerging markets} 

It is important to note that for any set of random variables, their copula function contains all information about the dependence structure between the random variables as well as between any rank-preserving transformations of the random variables \cite[see, e.g.,][for a survey of copula methods in finance]{cherubini/etal:11}.  Therefore, if we use the ranks of the original data instead of the original data itself, the structure of information transfer will remain unchanged. Meanwhile, the estimation is now over ranks and $H$ uses conditional copula density, rather than conditional density. 
This makes heavy tails in the marginal distributions innocuous and focuses the procedure on shifts in the conditional distributions when the first two moments of the data cannot be estimated as would be the case during extreme market conditions such as financial crises and contagion episodes.  

Capturing tail-dependence in financial data is another attractive feature of copula-based modelling. Tail-dependence refers to the phenomenon where extreme movements in one financial market are associated with extreme movements in another, particularly during market downturns. In emerging economies, establishing tail-dependence is crucial to effective risk management and portfolio diversification, to portfolio managers and international supervisory authorities. 

There is an increasing volume of evidence for tail-dependence among markets in emerging economies. For instance, \cite{Tian:23} find evidence of asymmetric negative and nonlinear tail-dependence between foreign exchange rates (FX) and stock markets for Brazil, Chile, Hungary, India, Mexico, Philippines, Poland, Russia, South Africa, and Thailand. \cite{RODRIGUEZ2007401} study financial contagion during the Asian and Mexican crises and find evidence of strong tail-dependence in Asia and symmetric dependence in Latin America. \cite{NAEEM2023100971} analyze monthly returns on MSCI emerging markets stocks indices for 23 countries from January 1995 to May 2021 and find evidence of strong tail-dependence across all the emerging markets during crises. 
Importantly, these and other studies highlight the limitations of traditional correlations as measures of information flow.

\section{Simulation experiments}\label{sec:sims}

\subsection{Simulation designs}\label{subsec:sim_design}

In this section, we compare performance of the baseline and extension methods with alternative procedures and among themselves. We use synthetic data which is representative of real-life market data from emerging economies in terms direction and intensity of information transfer. We use two common scenarios from information transmission literature, namely, termination of transfer and inversion of transfer. The two scenarios correspond to what is known as sudden stops \cite[see, e.g.,][]{Hassan:24} and spillover shift \cite[see, e.g.,][]{AuYong/etal:04}. The sample size we consider is $T=1000$.

(a) Termination of the information transfer. This is when the information transfer from $X$ and $Z$ to $Y$ stops at the change-point. We model this using the following data generating process (DGP): 
\begin{equation*} Y_t = 
    \begin{cases}
      0.5 \ln X_{t-1} + 0.2 Z_{t-1}^2 + 0.3 \nu_t, \hspace{0.3cm} t \leq 500\\
      \xi_t, \hspace{0.3cm} t > 500
    \end{cases}\,
\end{equation*}
where $\nu_1, ... , \nu_{500} \overset{\mathrm{i.i.d.}}{\sim} \mathcal{N}(0,1)$, $\xi_{501}, ... , \xi_T \overset{\mathrm{i.i.d.}}{\sim} \mathcal{N}(5,36)$, $X_1, ... , X_T  \overset{\mathrm{i.i.d.}}{\sim} Exp(3)$, and $Z_1, ... , Z_T  \overset{\mathrm{i.i.d.}}{\sim} \Gamma(3,1)$. We use the exponential and gamma distributions as well as the power and log transformations to permit a non-normal distribution of $Y_t$ and nonlinear patterns of information transfer.  

This scenario is particularly relevant in modelling a sudden stop in international credit flows to emerging economics, which  may bring about financial and balance of payments crises \citep{Calvo:98}. Sudden stops were first identified with reference to the oil shocks in the late 1970s. They were detected in the 1990s following the Mexican crisis (1994), Argentina's external debt crisis (1995), and the Asian (1997) and the Russian (1998) financial crises. More recent examples of sudden stops triggered by increases in macroeconomic uncertainty were observed during the COVID-19 pandemic. Increased political uncertainty and sanctions caused sudden stops in Russia and Ukraine during the 2014 annexation of Crimea and the 2022 start of military conflict between the two countries \cite[see, e.g.,][]{KONYA2024103110}.  

Sudden stops can take the form of large and unexpected reversals in foreign capital inflows \cite[see, e.g.,][]{calvo2004empirics}. This justifies a separate analysis of the second scenario. 

(b) Inversion of the information transfer. This is when an information flow from $X$ and $Z$ to $Y$ is reversed at a change-point:
\begin{equation*} Y_t = 
    \begin{cases}
      0.5 \ln X_{t-1} + 0.2Z_{t-1}^2 + 0.3 \nu_t, \hspace{0.3cm} t \leq 500\\
      \xi_t, \hspace{0.3cm} t > 500
    \end{cases}\,
\end{equation*}
where 
\begin{equation*} X_t = 
    \begin{cases}
      \upsilon_t, \hspace{0.3cm} t \leq 500\\
      0.3Y_{t-1} + 0.1 \psi_t, \hspace{0.3cm} t > 500
    \end{cases}\,
\end{equation*}
\begin{equation*} Z_t = 
    \begin{cases}
      \delta_t, \hspace{0.3cm} t \leq 500\\
      0.6Y_{t-1}^2 + 0.1 \zeta_t, \hspace{0.3cm} t > 500
    \end{cases}\,
\end{equation*}
and where $\nu_1, ... , \nu_{500}, \psi_{501}, ..., \psi_{T}, \zeta_{501}, ..., \zeta_{T} \overset{\mathrm{i.i.d.}}{\sim} \mathcal{N}(0,1)$, $\xi_{501}, ... , \xi_{T} \overset{\mathrm{i.i.d.}}{\sim} \mathcal{N}(5,9)$ and $    \upsilon_1, ... , \upsilon_{500}  \overset{\mathrm{i.i.d.}}{\sim} Exp(3)
$, $\delta_1, ... , \delta_{500}  \overset{\mathrm{i.i.d.}}{\sim} \Gamma(3,1)$. In this scenario, the information flow switches direction in the middle of the sample and starts flowing from $Y$ to $X$ and $Z$, not from $X$ and $Z$ to $Y$. 

This scenario is relevant when capital flows in emerging markets are not only reduced but also reversed, resulting in net outflows of capital. These reversals are often more severe than sudden stops, as they involve both a decline in inflows and an increase in outflows. Reversals are frequently linked to domestic factors, such as economic imbalances, political instability, or financial crises \cite[see, e.g.,][]{zhang/etal:24}.

The summary statistics of the simulated data using the two DGPs are presented in Table \ref{tab:sum_stats1}. The data are leptokurtic, asymmetric and weakly dependent both in levels and in ranks, which are typical characteristics of financial and economic data measured in log-return or growth-rate forms \cite[see, e.g.,][]{cont:01}. 

\begin{table}[H]
    \centering
    \begin{tabular}{c|cccccc} 
         regime& mean& std&  skewness&  kurtosis&  corr$(Y_t; Y_{t-1})$& rank corr\\ \hline 
         \multicolumn{5}{c}{Information transfer termination}\\ \hline 
         before change& 2.541& 2.768& -2.106&  5.812&  -0.062& -0.040\\ 
         after change& 5.126& 5.968& 0.110&  0.241&  -0.076& -0.080\\
         \hline 
         \multicolumn{5}{c}{Information transfer inversion}\\ 
         \hline 
         before change& 2.541& 2.768&  -1.527&  2.973&  0.040& -0.040\\ 
         after change& 5.126& 2.99& -0.192& 0.137& 0.007&  -0.064\\ 
        \end{tabular} \caption{Summary statistics of synthetic data}
\label{tab:sum_stats1}
\end{table}

\subsection{Comparison with traditional procedures}

We start by benchmarking the new approach against the conventional applications of the SR procedure, that is, without the use of entropy. The conventional approaches we consider are as follows: (a) application of the SR statistic to detect a change in the mean of the target variable $Y_t$ directly; (b)
application of the SR statistic to detect a change in the mean of the residuals ${e}_t$ from the AR model in Eq.~(\ref{eq:resid}); (c) application of the SR statistic to detect a change in the mean of the fitted values $\hat{e}_t$ from Eq.~(\ref{eq:granger}). We focus on termination of information  transfer.\footnote{The inversion of information transfer scenario is available upon request.} 
The SR window length is $\Delta=50$.

\begin{figure}[H]
     \begin{subfigure}[b]{0.5\textwidth}
         \centering
         \includegraphics[width=\textwidth]{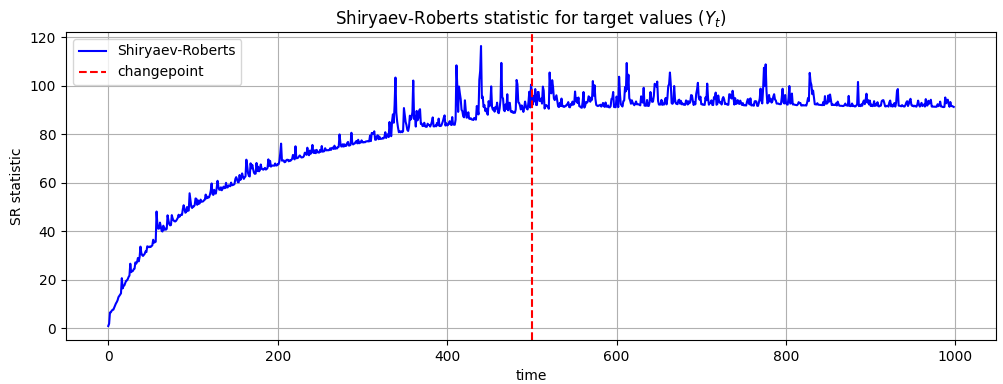}
     \end{subfigure}
     \hfill
     \begin{subfigure}[b]{0.5\textwidth}
         \centering
         \includegraphics[width=\textwidth]{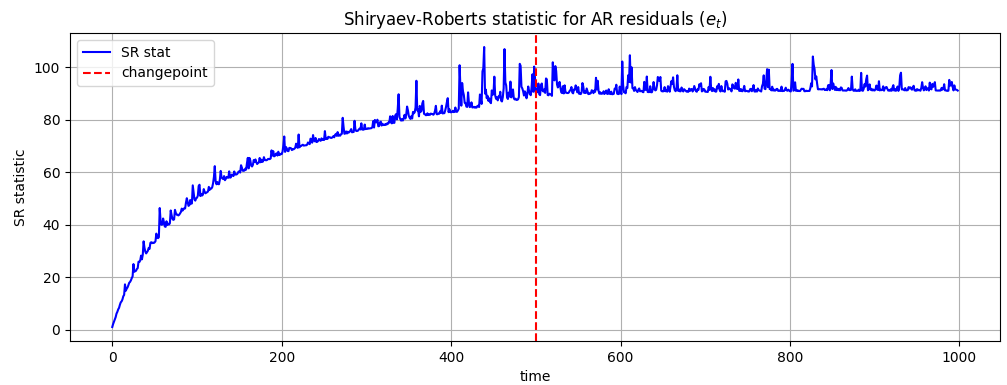}

     \end{subfigure}
     \vfill
     \begin{subfigure}[b]{0.5\textwidth}
         \centering
         \includegraphics[width=\textwidth]{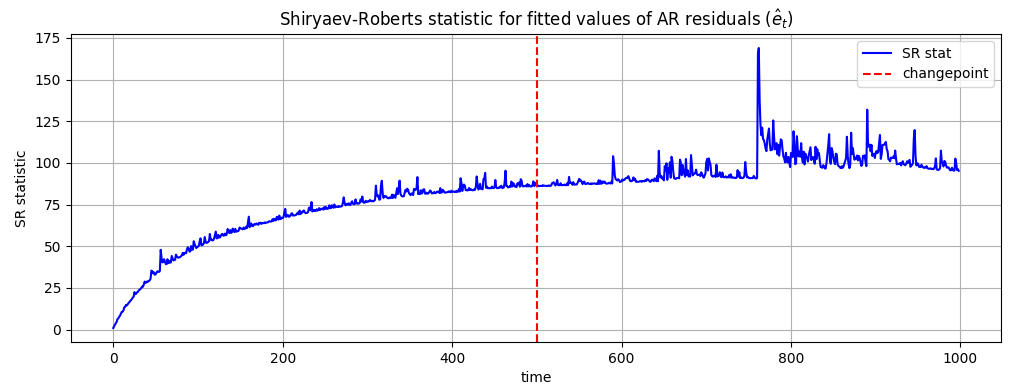}
     \end{subfigure}

    \hfill
     \begin{subfigure}[b]{0.5\textwidth}
         \centering
         \includegraphics[width=\textwidth]{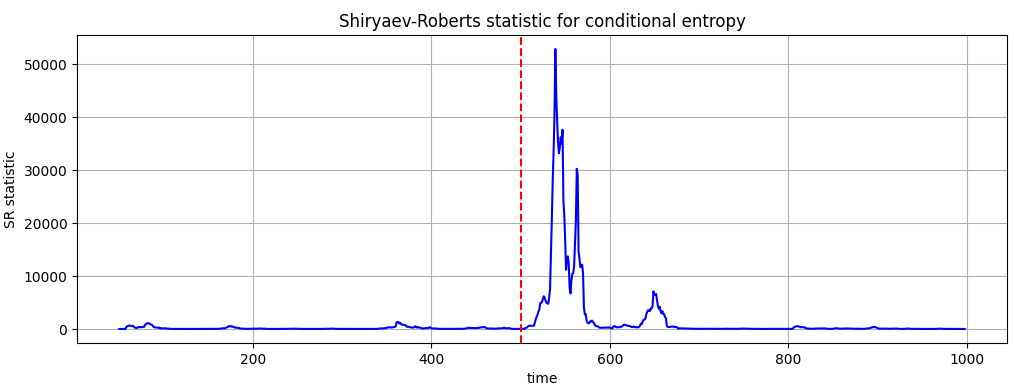}
     \end{subfigure}
        \caption{Shiryaev-Roberts statistics for termination of information transfer}
        \label{fig:SR_traditional}
\end{figure}

Figure \ref{fig:SR_traditional} shows the values of respective SR statistics and the location of the change-point (vertical dashed line). It is clear from the figure that the traditional ways of applying the SR procedure fail to detect the change-point. The SR statistics for three of the four figures grows with time and flattens out near the change-point but no spikes are large enough to trigger an alarm. On the other hand, the entropy-based procedure shown on the lower right panel produces an immediate and strong alarm. It is remarkable that the entropy-based SR statistic does not follow the same pattern as the conventional versions. 

This difference in performance is due to the limitation of traditional approaches. When applied to specific variables, as opposed to the entropy, they seem to target shifts of specific nature, e.g., shifts in mean, variance, autocorrelation. In this case, the shift in mean is too small. A key advantage of our approach is that it captures jumps in \emph{information transfer}. In other words, instead of analyzing various shifts in SR statistics, each detecting a specific drift, the entropy-based measure directly addresses the stability of the entire relationship. 

In the setting of sudden stop detection in emerging markets, this allows the researcher to create a more comprehensive and conservative early warning system, sensitive to a wider range of triggers and more complex ways they affect the distribution of the target variable.

\subsection{Nonlinearity and ranks}

\begin{figure}[H]
     \begin{subfigure}[b]{0.5\textwidth}
         \centering
         \includegraphics[width=\textwidth]{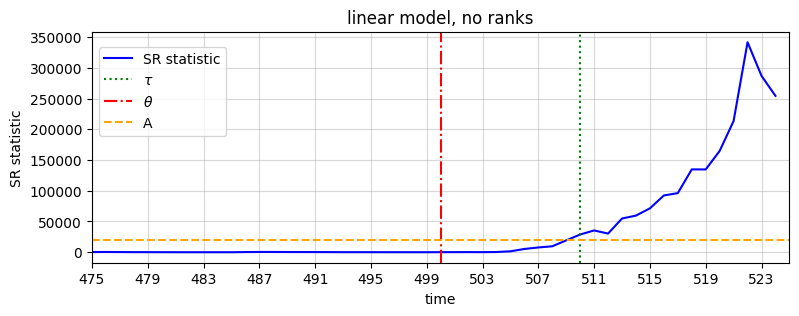}
     \end{subfigure}
     \hfill
     \begin{subfigure}[b]{0.5\textwidth}
         \centering
         \includegraphics[width=\textwidth]{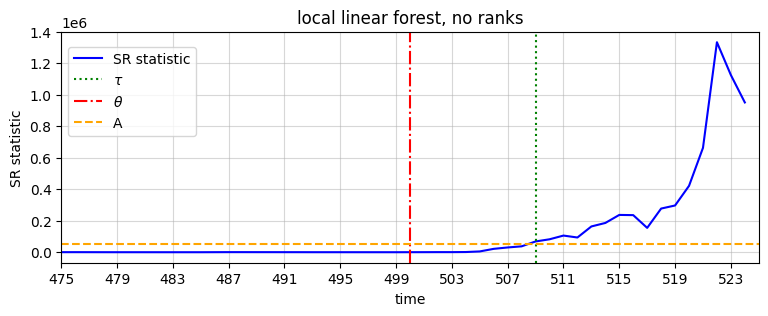}
     \end{subfigure}
     \vfill
     \begin{subfigure}[b]{0.5\textwidth}
         \centering
         \includegraphics[width=\textwidth]{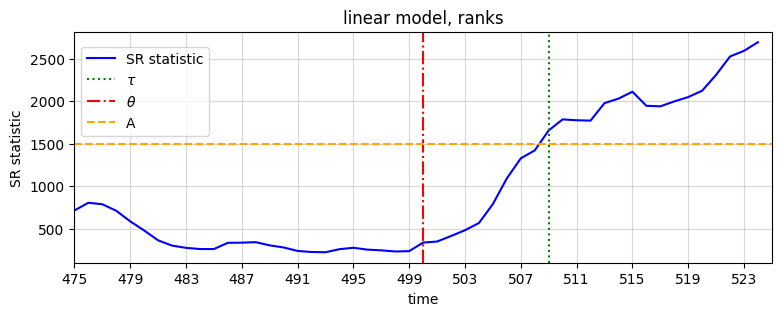}
     \end{subfigure}
    \hfill
     \begin{subfigure}[b]{0.5\textwidth}
         \centering
         \includegraphics[width=\textwidth]{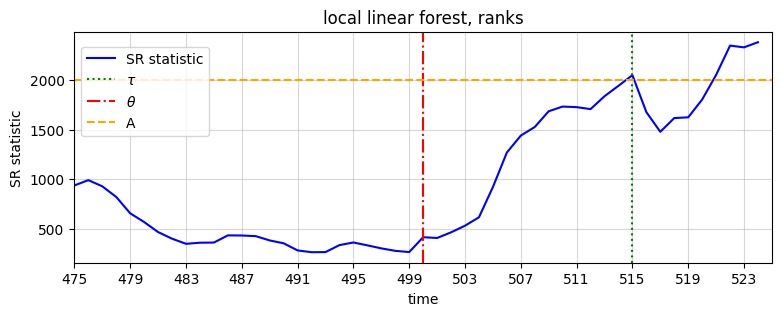}     \end{subfigure}
        \caption{Shiryaev-Roberts statistic for termination of information transfer}
        \label{fig:SR_termination}
\end{figure}

\begin{figure}[H]
     \begin{subfigure}[b]{0.5\textwidth}
         \centering
         \includegraphics[width=\textwidth]{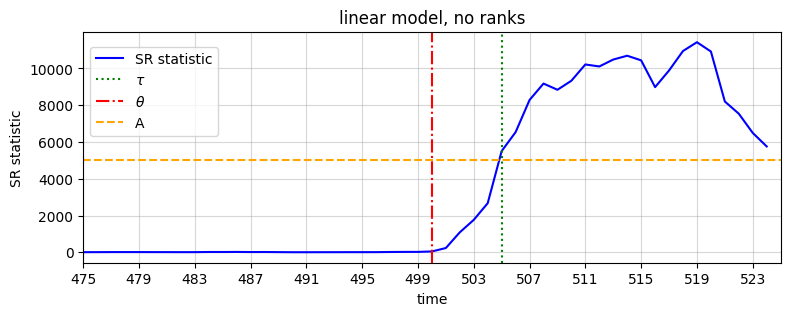}
     \end{subfigure}
     \hfill
     \begin{subfigure}[b]{0.5\textwidth}
         \centering
         \includegraphics[width=\textwidth]{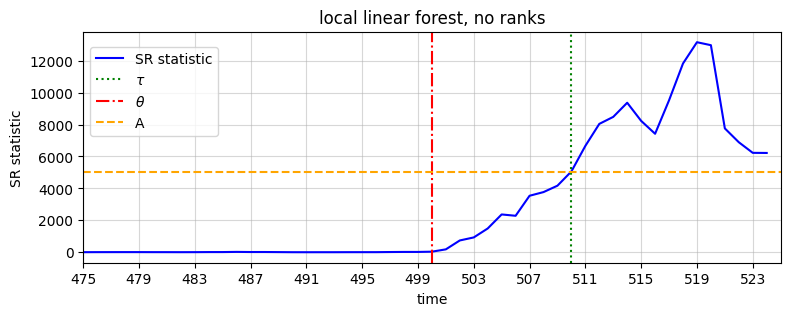}

     \end{subfigure}
     \vfill
     \begin{subfigure}[b]{0.5\textwidth}
         \centering
         \includegraphics[width=\textwidth]{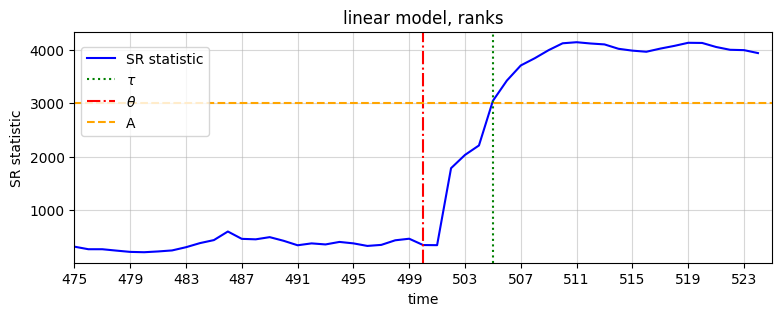}
     \end{subfigure}
    \hfill
     \begin{subfigure}[b]{0.5\textwidth}
         \centering
         \includegraphics[width=\textwidth]{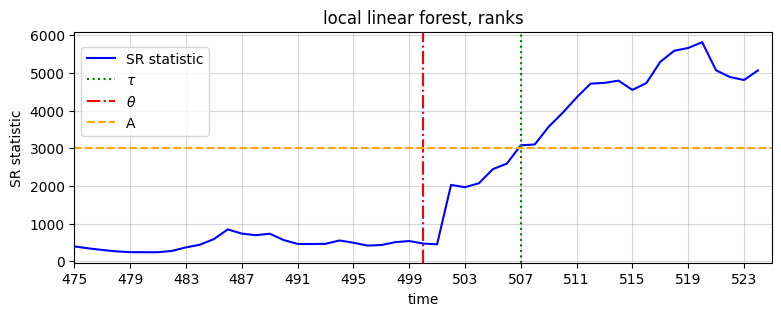}
     \end{subfigure}
        \caption{Shiryaev-Roberts statistic for inversion of information transfer}
        \label{fig:SR_inversion}
\end{figure}

We now turn to the detection of transfer termination and inversion using the proposed extensions of the baseline entropy approach, that is, using local linear forests and ranks. 
That is, we use the two scenarios (termination and inversion) to simulate the target variable $Y_t$ as described in Section \ref{subsec:sim_design}, and apply four ways of computing the SR statistic: (a) the baseline version which uses no random forest and no ranks; (b) the extension with LLF but no ranks; (c) the extension with ranks of $\hat{e}_t$ but without random forests; and (d) the extension with ranks of $\tilde{e}_t$, which uses local linear forest and ranks.  Our simulation results are summarized in Figures (\ref{fig:SR_termination})-(\ref{fig:SR_inversion}), where $\theta$ denotes the moment of change, $\tau$ denotes the moment of change detectionm and $A$ is the threshold value, used for the Shiryaev-Roberts statistic. The window length is $\Delta=50$. 

It follows from Figures (\ref{fig:SR_termination})-(\ref{fig:SR_inversion}) that all extensions of the baseline method work very well to detect both terminations and reversals. There is no clear order in the performance among the four methods. For both termination and inversion, all four methods are able to produce an alarm in 7-15 time periods, using the respective thresholds; Table \ref{tab:detection_delay} summarizes this finding.

\begin{table}[H]
    \centering
     \begin{tabular}{c | c c} 
     Model & Termination & Inversion  \\ [0.5ex] 
     \hline
     linear & 10 & 5 \\ 
     LLF & 9 & 10  \\
     linear + ranks & 9 & 5  \\
     LLF + ranks & 15 & 7  \\
     \end{tabular}
     \caption{Change-point detection delay $(\tau - \theta)$}
     \label{tab:detection_delay}
\end{table}

The similar performance of the new methods in detecting both sudden stops and reversals makes them attractive for determining whether some types of capital flows in emerging markets are more likely to reverse than others. \cite{SULA2009296} point out that earlier statistical tests have yielded conflicting results on this issue and find evidence that direct investment is less reversible but private loans and portfolio flows are more reversible in a sample of 35 emerging economies. 

Interestingly, a comparison between the lower and upper panels of the figures 
suggests that PFA for ranks is higher than for levels, as can be seen by the non-zero values of SR before the break in the lower panels (and zero SR in the upper panels). This could be interpreted as suggesting that using ranks is less reliable than using levels. It is worth noting that ranks make the procedure robust to undesirable features of the marginal distributions such as heavy tails, but they also lose the information contained in the marginals, making the resulting procedure less powerful when that information is useful. 

This suggests that in empirical studies in which heavy tails and tail-dependence are not of concern, the researcher may prefer using the original data rather than the ranks. However, as we argued earlier, emerging markets are characterized by unusually heavy tails and strong dependence in extremes, that cannot be adequately captured by conventional measures. Moreover, sudden stop and reversals are usually accompanied by crises and contagions. Therefore, it is important to confirm that the proposed rank-based methods are effective in such a circumstance. 

\subsection{Heavy tails and tail-dependence}

To further investigate the benefits of rank-based measures we consider a setting where the heavy-tailed property of the individual distributions prevents us from using conventional measures such as the mean, variance, or correlations, to detect sudden stops and reversals. This setting is of particular relevance for the analysis of financial data from emerging markets, especially during crises \cite[see, e.g.,][]{CHAUDHURI:03, Chen/ibra:19}. 

We modify the simulation design of the previous sections by including a component with heavy-tailed data and tail-dependence.  For this purpose we add the penalty $\lambda$ to a simplified version of the DGP, which represents termination of the information flow:
\begin{equation*} Y_t = 
    \begin{cases}
      X_{t-1} + (1 -\sqrt{\lambda_{t-1}})\xi_t, \hspace{0.3cm} t \leq 500,\\
      0.9 + \xi_t, \hspace{0.3cm} t > 500,
    \end{cases}\
\end{equation*}
where $\xi_1, ..., \xi_T \overset{\mathrm{i.i.d.}}{\sim} \mathcal{N}(0, 0.5), X_1, ..., X_{500} \overset{\mathrm{i.i.d.}}{\sim} WeibMin(1.5)$. 
The role of $\lambda$ is to control the noise level depending on how close the observation is to the sample median. Let $d_t = X_t - \text{med}(X_t)$ denote such a distance. We define $\lambda$ as follows: 
$$ \lambda_t = \dfrac{d_t - \min_{t}\{d_t\}}{\max_t\{d_t\} - \min_{t}\{d_t\}} \in [0,1].$$

Clearly, the closer an observation is to the median, the more noise  $Y_t$ has over $X_{t-1}$. Observations that lie closer to the sample extremes have $Y_t$ that are closer to being perfectly collinear with $X_{t-1}$. This produces non-zero tail-dependence between $X_{t-1}$ and $Y_t$, that is, a non-zero correlation between $X_t$ and $Y_t$ when they get closer to the extremes of their distributions.  Additionally, we use the Weibull minimum extreme value distribution for $X_t$, rather than normal or exponential, to induce the heavy-tailed property in the marginals.

\begin{figure}[H]
     \begin{subfigure}[b]{0.5\textwidth}
         \centering
         \includegraphics[width=\textwidth]{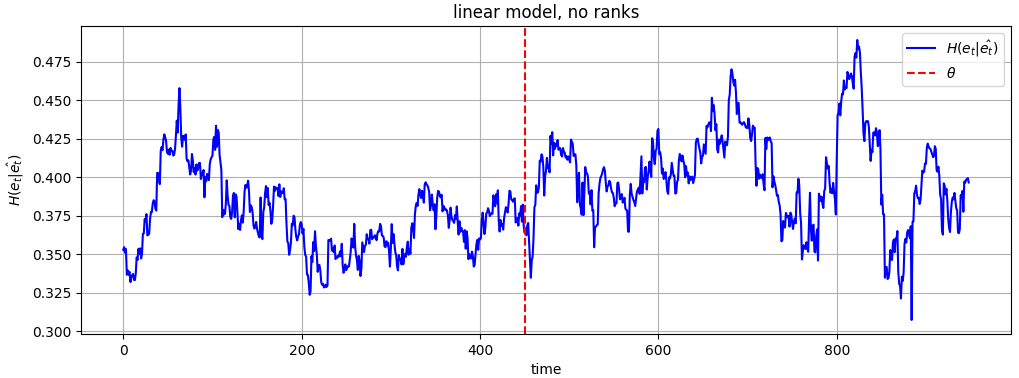}
     \end{subfigure}
     \hfill
     \begin{subfigure}[b]{0.5\textwidth}
         \centering
         \includegraphics[width=\textwidth]{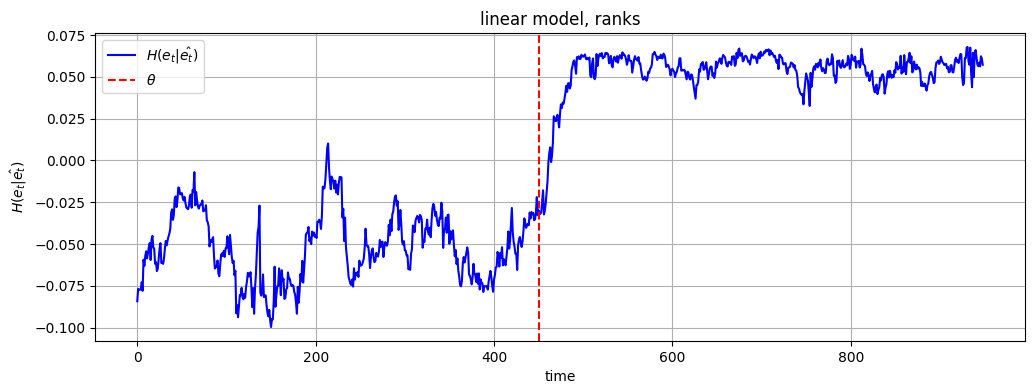}

     \end{subfigure}
     \vfill
     \begin{subfigure}[b]{0.5\textwidth}
         \centering
         \includegraphics[width=\textwidth]{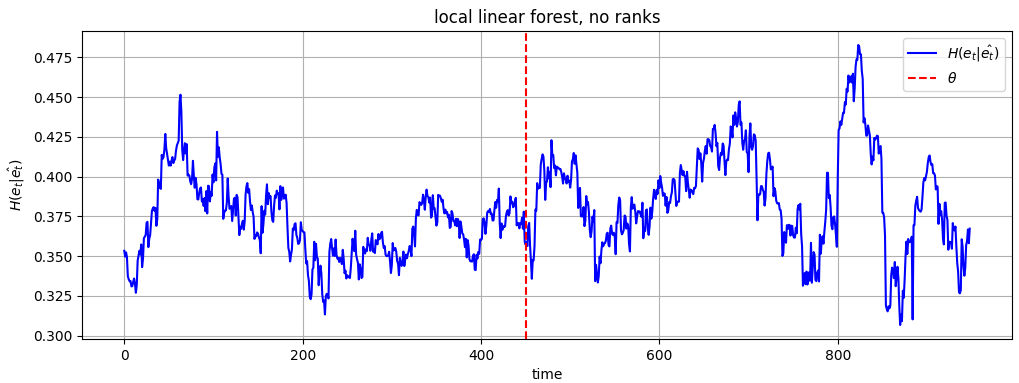}
     \end{subfigure}
    \hfill
     \begin{subfigure}[b]{0.5\textwidth}
         \centering
         \includegraphics[width=\textwidth]{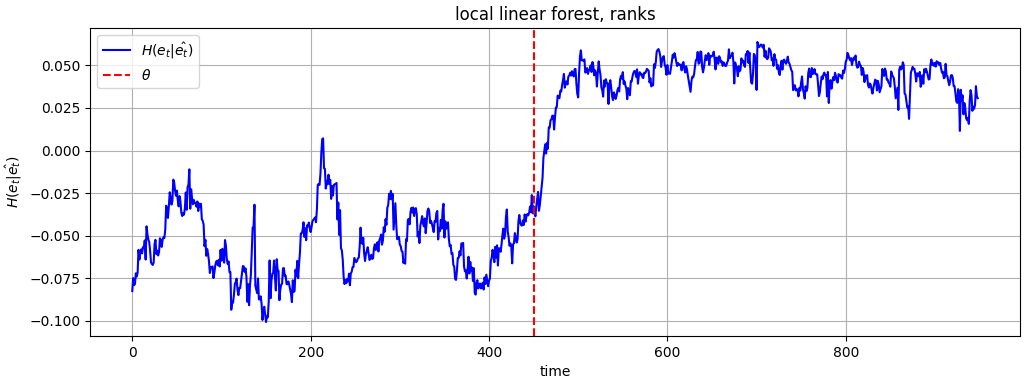}     \end{subfigure}
        \caption{Estimates of entropy for heavy-tailed data with tail-dependence}
        \label{fig:entropy_heavy_tailed)}
\end{figure}

Figure \ref{fig:entropy_heavy_tailed)} plots the conditional entropy processes $H$ obtained using the four methods we consider: (a) the baseline method that does not use ranks or random forest, (b) the methods that uses local linear forest but without ranks, (c) the method that uses ranks but without random forest, (d) the method that uses local linear forest and ranks. The vertical dashed line indicates the change-point $\theta$. We do not compute the SR statistics because this is unnecessary for the point we are making. 

It is evident from Figure \ref{fig:entropy_heavy_tailed)} even without computing the SR statistic that the entropy-based approach using the original data fails to detect the regime shift (the upper and lower left panels do not show a visually significant change in $H$), while the two rank-based methods provide consistent results and rapidly detect the regime shift (the upper and lower right panels show a clear jump to a new mean of $H$ after the change-point). This suggests that the new rank-based methodology is effective in cases when data are heavily leptokurtic and tail-dependent.\footnote{The results for information reversal are similar and are not reported to save space.} 

\subsection{Comparison with other detection methods}

As an additional robustness check, we compare the proposed methods with non-SR approaches to concept drift detection from computer science such as Early Drift Detection Method (EDDM) and Adaptive Windowing Method (ADWIN) of \cite{Bifet}, and Kolmogorov-Smirnov Windowing Method (KSWIN) of \cite{Raab} and from economics such as an online version of the structural break detection method of \cite{Bai-Perron}.  We focus on information transfer termination as modelled in Section \ref{subsec:sim_design}. 

Since EDDM, ADWIN and KSWIN are designed to operate on univariate time series they need to be adapted to be compatible with the methods we propose. Similarly, the Bai-Perron test, being an offline test, needs to be turned into an online procedure. A perfectly fair comparison is impossible here as the original procedure were developed under somewhat different assumptions. The compromises we adopt are as follows:

(a) For ADWIN, EDDM, and KSWIN, we split time series $Y_t$ into non-overlapping segments of length 50, i.e., $[0,50], [51,100], \ldots.$ For each segment we obtain residuals $e_t - \hat{e}_t$, where $e_t$ and $\hat{e}_t$ are computed according to Eqs.~(\ref{eq:resid}) and (\ref{eq:granger}), respectively. We then concatenate these segments and apply the relevant algorithms to it. We repeat this 100 times and compute PFA (probability of false alarm), ADD (average detection delay) and ND (fraction of non-detections) over the replications.

(b) For the Bai-Perron test, we apply the test at the 1\% significance level to an expanding window of data starting with $[0, 400]$ and expanding to $[0, 401], [0, 402], \ldots.$ We stop when the test rejects the null. We repeat this 100 times and compute PFA, ADD and ND over the replications.

(c) For our proposed methods, we apply them as described in Section \ref{sec:proc} (baseline) and  \ref{subsec:llf} (LLF), with $\Delta = 50$, $\alpha = 0.5, \beta=0.9, m = 1$. We repeat this 100 times  and compute PFA, ADD and ND over the replications.








\begin{table}[H]
    \centering
     \begin{tabular}{c | c c c c c c} 
         Metric & ADWIN & EDDM & KSWIN & Bai-Perron & baseline & LLF\\ [0.2ex] 
     \hline
     
     PFA &  - & - & 0.89 & 0.66 & 0.1 & 0.09 \\ 
     ADD & - & - & 8.5 & 23.6 & 10.1 & 10.4 \\
     ND &  1 & 1 & 0.09 & 0 & 0.05 & 0.04  \\
     \hline
     \end{tabular}
     \caption{Performance comparison with non-SR methods}
     \label{tab:comparison}
\end{table}

The results are summarized in Table \ref{tab:comparison}. Using this simulation design, both ADWIN and EDDM failed to detect the break, while KSWIN detected it in 91\% of replications and produced a false alarm in $89\%$ of replications. This implies that there are only $2\%$ of replications where KSWIN detected the drift correctly. The Bai-Perron test detected all drifts, which is perhaps expected given the repeated testing problem implied by algorithm (b). It also produced 66\% false positives. It is therefore clear from the table that the proposed methods, both the baseline and LLF extension, strongly dominate the alternatives in terms of PFA, ADD and, aside from Bai-Perron, in terms of ND.

\section{Real data examples}\label{sec:appl}

In this section, we provide two empirical illustrations of how to apply our approach to create an early warning system in emerging markets.  

\subsection{Russia: market uncertainty and the conflict in Ukraine}

Many recent studies have found a significant impact of the Russian-Ukrainian conflict on global financial markets \cite[see, e.g.,][]{Izzeldin}. Less is known about information transmission in the time before the conflict broke out in February 2022. In this subsection, we explore whether the Russian stock market showed early warning signals that could have been detected using our approach. 

We consider the level of uncertainty captured by the conditional entropy -- and the corresponding SR statistic -- of RTSI, the main index of the Russian equity market, which is a capitalization-weighted average of the 50 most liquid Russian stocks traded on the Moscow Stock Exchange. The RTSI is an U.S. dollar-nominated index, known to attract foreign investors. 

The entropy is conditional on the information contained in key indicators of various international markets (commodity, equity, fixed-income) as well as key internal indicators. The list of variables is typical for models studying volatility spillover across financial markets of emerging economies and includes a stock market index of the emerging markets, a stock market index of a mature economy, an export-oriented exchange rate, a risk free rate and a government bond spread as well as an oil price and a few key domestic stocks dominated by internal investors \cite[see, e.g.,][]{Assaf, Sio-Chong}. The list is inevitably incomplete but sufficient to illustrate the main points. We provide details of the conditioning variables, their sources and summary statistics in Appendix B.  

We use a 100-day sliding window to compute the conditional entropy process $\hat{H}$ as described in Section \ref{subsec:llf}, that is, we use the local linear forest in the estimation of the conditional mean $\tilde{e}$ to allow for nonlinearity and to mitigate the curse of dimensionality. The order of the autoregressions in Eq.~(\ref{eq:granger}) is picked using BIC and is allowed to vary for each sliding window. We conducted sensitivity analysis, running the procedure for different values of the autoregressive order between 1 and 10 with no significance difference in the results. 

The number of shifts $m=100$, the exponential smoothing parameters are $\alpha=0.95$ and $\beta=0.95$. The period for which we compute $\hat{H}$ is May 27, 2021 to April 12, 2022. It contains 273 daily observations prior to the start of the military conflict on February 24,  2022 and 47 observations thereafter.

\begin{figure}[H]
     \begin{subfigure}[b]{\textwidth}
         \centering
     \includegraphics[scale = 0.65]{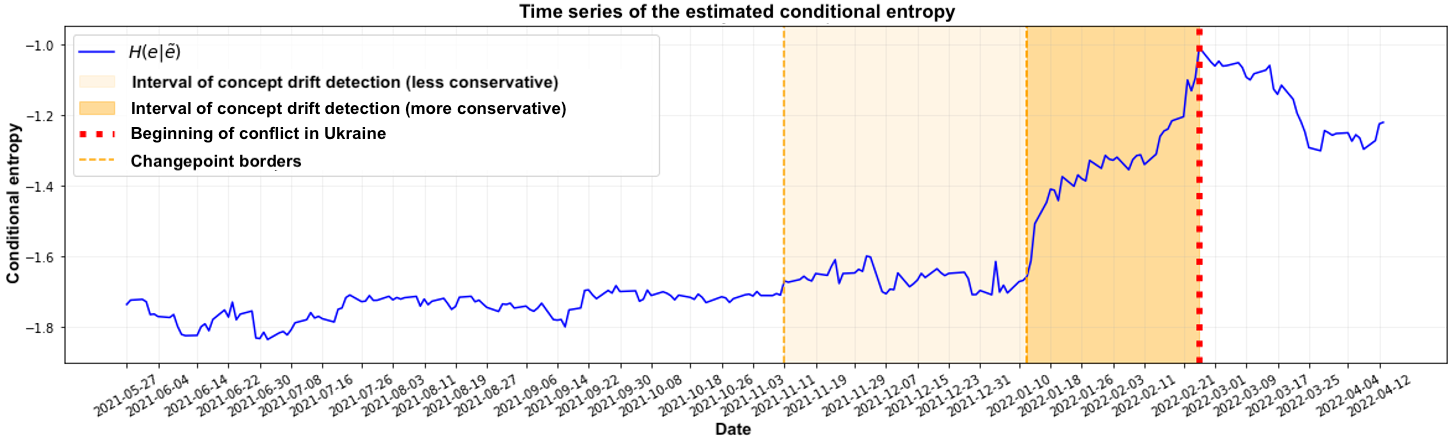}
     \end{subfigure}
     \vfill
     \begin{subfigure}[b]{\textwidth}
         \centering
 \includegraphics[scale = 0.65]{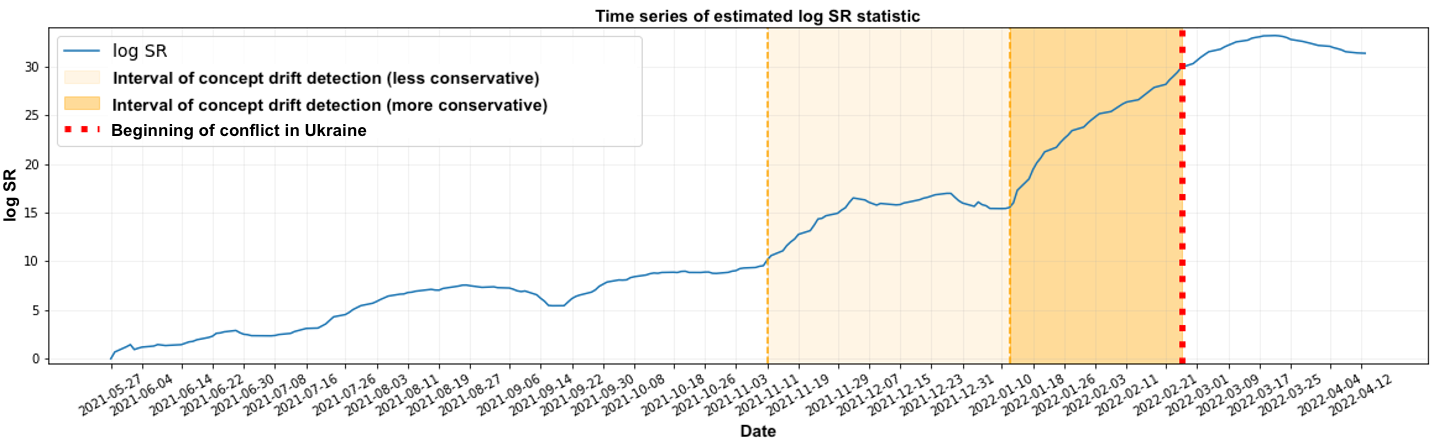}    
     \end{subfigure}
     \caption{RTSI-based conditional entropy and log SR statistic}\label{fig:entropy_RUS}
\end{figure}

Figure \ref{fig:entropy_RUS} plots the estimated conditional entropy and the corresponding SR statistic (on the log scale). Several observations from the figure are noteworthy. First, the conditional entropy reaches its peak precisely on February 24. Such precision is perhaps surprising, especially since the entropy measures the degree of disorder in RTSI after accounting for the information in a fairly rich conditioning set. Second, there is a rising trend in the SR statistic throughout the period, indicating a general increase in uncertainty. Third, the proposed method detects two change-points where the information flow changed significantly, at three months and at one month before February 24. Given the upward trend in SR in this example, we can think of the two change-points as specifying two levels of alarm intensity, more  and less conservative. 
We stress that nothing in the procedure targets the detection of the start of the conflict. In fact, there are no alarms following the second detected change-point and that period includes 
the first wave of sanctions, interest rate hikes, and capital controls, all of which followed much later than Februry 24. Yet, the RTSI entropy captures the accumulated market uncertainty of the preceding period, with the peak on February 24 and two warnings, triggered at least a quarter earlier. 

As an additional insight, it is worth noting that the October 1, 2022 alert corresponds to a significant fall in the Russian stock and currency markets, following the announcement of new sanctions on Russia. Interestingly, after a large drop in RTSI (of about 20\% in 12 days), the Russian bond market showed an even more rapid growth (13.6\% in 6 days). This can be viewed as additional empirical evidence supporting the pattern established by the literature on financial crises that credit market booms serve to counteract or fuel asset market busts \cite[see, e.g.,][]{Greenwood/etal:22, Jorda/etal:15}.

This empirical example also highlights the importance of volatility jumps in generating early warnings. 
The entropy and SR-statistic increases coincide with significant growth of volatility during this period, a well established precursor of crises 
\cite[see, e.g.,][]{Kharroubi, Xin, Laborda}.

\subsection{South Africa: financial spillover during COVID-19 pandemic}

Global financial spillovers that happened around the time of the COVID-19 pandemic have been widely documented \cite[see, e.g,][]{McKibbin, Gagnon}. With regard to the BRICS economies,  \cite{KHALFAOUI/etal:23} explore the spillover patterns  using a large number of financial indicators, including market indices, exchange rates, cryptocurrencies prices, evaluated at several quantiles of their distributions. As a result, 
\cite{KHALFAOUI/etal:23} offer a number of quantile- and country-specific directional networks reflecting the interconnectedness between conventional and Islamic BRICS stock markets, crypto markets (Bitcoin, Ethereum, Litecoin) and commodity markets (oil and gold). In this subsection, we draw on their results to further study information transfer around the time of the pandemic affecting the South African financial market.  

Specifically, when looking at the middle quantile, \cite{KHALFAOUI/etal:23} find evidence of information transfer  between VIX (global volatility index), OVX (oil volatility), GVZ (gold volatility), cryptos (BTC, ETH, LTC), as well as conventional and Islamic indexes; see Appendix B for details and sources of the data. These connections become stronger at extreme quantiles and may change direction. For example, they note that MXSAI (MSCI Islamic South Africa), MXSA (MSCI South Africa), and TEU (Twitter Economic Uncertainty) transmitted stronger shocks to VIX when the markets were bearish or bullish than when they were normal. 

We expand on this result by applying one of the proposed change-point detection method to South Africa's MXSA index, an equity index measuring the performance of 32 large and mid-cap South African stocks. Its behavior conditional on other indicators reflects spillover effects in the period preceding the COVID-19 pandemic. We evaluate the MXSA entropy using the methods of Section \ref{subsec:llf}, conditional on the variables that were found to be important transmitters by \citet[][Figure 7]{KHALFAOUI/etal:23}. 
The data covers the period June 2, 2019 to May 28, 2021.\footnote{We are grateful to Professor Rabeh Khalfaoui for providing their data to us.} The exponential smoothing parameters are $\alpha=0.75$ and $\beta=0.75$, the 
number of shifts $m=100$ and the size of sliding window is 50 observations.

\begin{figure}[H]
     \begin{subfigure}[b]{\textwidth}
         \centering
     \includegraphics[scale = 0.4]{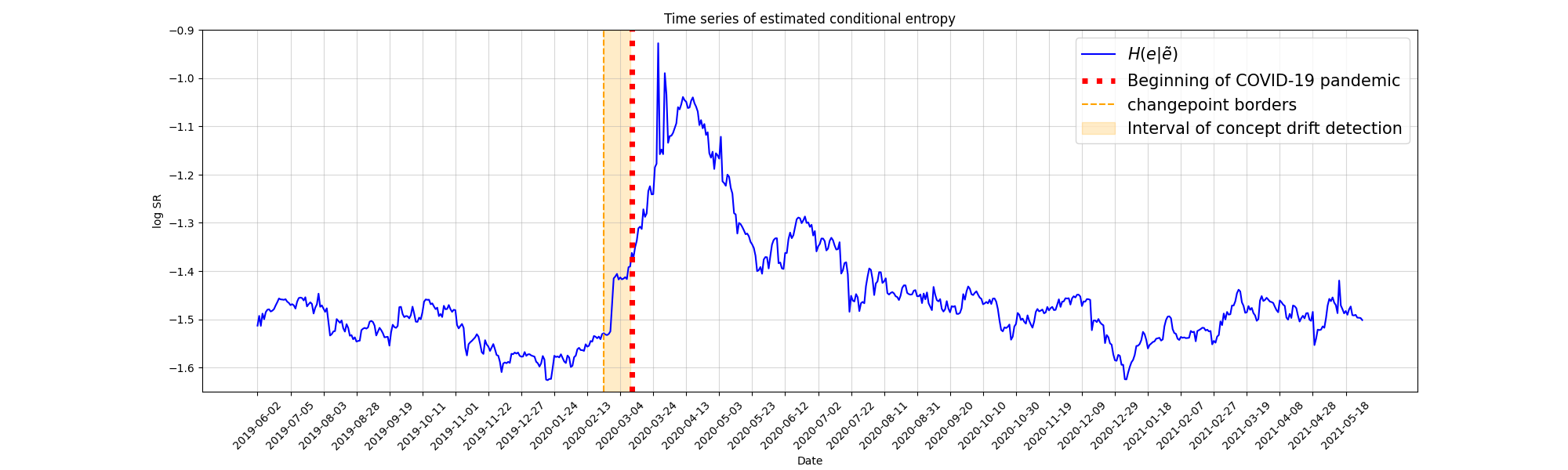}
     \end{subfigure}
     \vfill
     \begin{subfigure}[b]{\textwidth}
         \centering
 \includegraphics[scale = 0.4]{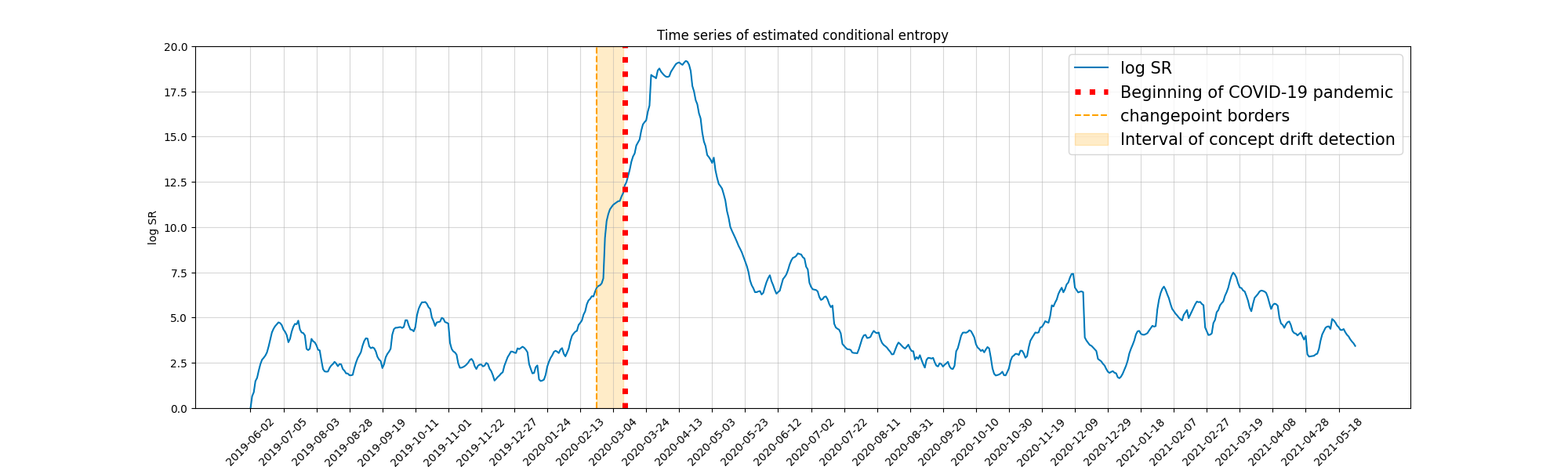}    
     \end{subfigure}
     \caption{MXSA-based conditional entropy and log SR statistic}\label{fig:entropy_SouthAfrica}
\end{figure}

Figure \ref{fig:entropy_SouthAfrica} shows the estimated conditional entropy and log SR statistic as well as a change-point detected on February 23, 2020 and the official start of the pandemic on March 11, 2020. Again, we observe a strong upward trend in the MXSA-based entropy as more uncertainty spills over into the South African stock market. The change-point can be detected about a month ahead of the highest entropy point, which is reached in late March - early April of 2020, and a couple of weeks ahead of the official start of the pandemic. This detection is a reflection of a concept drift that may have started with the first news about COVID in the late 2019. 

The patterns displayed on the figure show  a rapid increase and a somewhat slower mean reversion after the onset of a recession in the mid and late 2020. The mean reversion is perhaps noteworthy. It suggests that our approach can be used for impulse response analysis -- something that we leave for future work. 


Interestingly, in this empirical example, 
the peak of entropy does not coincide with the official beginning of the pandemic. The official announcement followed about 16 days after the alarm and the entropy was increasing for over a month after that. In April, the South African government started introducing restrictions and policies that must have contributed to the slow reduction in entropy. 

At the time of the alarm, there was an increase in volatility in the market, combined with a drop in conventional assets prices (shares, bonds, commodities, currencies) and a rapid growth 
of major cryptocurrency prices 
(BTC, ETH) in the short-run. This can provide further empirical support for the link between the information about COVID-19 health outcomes and the crypto prices uncovered by \cite{Sarkodie}.


\section{Concluding remarks}\label{sec:concl}

We developed a new framework for the construction of early warning systems, designed to detect change-points via conditional entropy estimation. This is an original approach that combines benefits of information theory and machine learning to help monitor the level of uncertainty, remaining after accounting for lags of the target and explanatory variables. This approach is generic but, as we show, it is particularly relevant for modeling information flows in emerging markets which are differentiated from other markets by the frequency of their structural break, tail-dependence, and heavy tails in the data. 

We propose a baseline method which uses a new conditional density estimator based on random forests, and we work out two extensions to the baseline that better capture nonlinear interactions and accommodate heavy-tailed distributions. The two extensions use a local linear forest estimator of the conditional mean and a rank-based estimator of the conditional copula density, respectively. 

The entropy-based test is designed to detect changes in the entire conditional distribution rather than in specific moments. As a result, it exhibits strong power against alternatives involving shifts in higher-order moments, nonlinear dependence, or tail behavior—features that are common in emerging market data. This is consistent with the simulation evidence in Sections 4.2–4.4, where the method successfully detects changes that are not visible in mean-based SR procedures.

At the same time, when the change is limited to a small shift in a single moment (e.g., a weak mean shift), specialized parametric tests may exhibit higher power. Furthermore, rank-based versions of the procedure improve robustness to heavy tails and tail dependence but may lose efficiency when information in the marginal distributions is informative. These trade-offs highlight that the proposed method is particularly well suited to environments where distributional changes are complex and not confined to low-order moments.

We demonstrate the use of the new framework on synthetic and real-life data. The simulation results show when the conventional SR statistic and alternative approaches from machine learning and economics fail while the proposed approach exhibits a robust performance. 
Fundamentally, the proposed methods constitute a first step towards leveraging recent advances in machine learning and statistics in automated market monitoring, used 
in markets with pronounced volatility, structural breaks, heavy tails and tail-dependence. We leave further extensions of the proposed framework, such as impulse response analysis using conditional entropy, for future work. 


\setstretch{1}

\bibliography{literature.bib}
\newpage
\appendix
\section*{APPENDIX}

\section{Change-point detection using entropy}\label{app:algo}

Algorithm \ref{alg:SR} contains a pseudo-code for the Shiryaev-Roberts procedure which produces change-points $\tau$ using entropy process $H$.

\textbf{Data:} Time series $H_1, H_2, ..., H_n$ as defined in Section \ref{sec:prelims}; set of possible shifts in mean $M_\Delta$ and threshold $A$, as described in Section \ref{sec:proc}.

\textbf{Starting values:} $SR_0 = 0; \hat{\mu}_1 = H_1; \hat{\sigma}_1^2 = 1$;

\begin{algorithm}[H]
    \caption{Shiryaev-Roberts procedure for change-point detection in entropy}\label{alg:SR}

    \While{$2\le t \leq n$}{
    $\xi_t = \dfrac{H_t - \hat{\mu}_{t-1}}{\hat{\sigma}_{t-1}}$\;
        \For{${\Delta}\mu_{i} \in M_{\Delta}$}{
          
          $SR_t^{(i)} = (1+SR_{t-1}^{(i)})\cdot \exp\{{\Delta}\mu_{i}(\xi_t - \frac{{\Delta}\mu_{i}}{2} )\}$\;
        }
        $SR_t^{w} = \dfrac{1}{m}\sum\limits_{i = 1}^{m} SR_t^{(i)}$

        \If{$SR_t^{w} > A $}{
            $\tau = t$\;
            \textbf{break}
            }

       $\hat{\mu}_t = \alpha \hat{\mu}_{t-1} + (1-\alpha)H_t$\;
       $\hat{\sigma}_t^2 = \beta \hat{\sigma}_{t-1}^2 + (1-\beta)(H_t - \hat{\mu}_t)^2$\;

        $t=t+1$;
    }
        \KwResult{change-point $\tau$}

    \end{algorithm}

\section{Data description}\label{app:sources}

\subsection{Variables and sources}

\textbf{Russia:}
    \begin{itemize}
        \item Russian Trading System (RTS) Index:
    \url{https://www.moex.com/en/index/RTSI/archive}
        
        \item S\&P500: \url{https://www.spglobal.com}
        
        \item USD/CNY exchange rate: \url{https://finance.yahoo.com/quote/CNY%3DX/history}
        
        \item Two-year Russian government bonds prices: \url{https://cbonds.ru/country/Russia_bond/}
        
        \item Sberbank shares: \url{https://www.tradingview.com/symbols/ALOR-SBER}
        
        \item Brent crude oil futures: \url{https://www.investing.com/commodities/brent-oil-historical-data}
        
        \item  Ruble Overnight Index Average (RUONIA): \url{https://www.cbr.ru/hd_base/ruonia/dynamics/}
        
        \item MSCI for emerging markets: \url{https://www.investing.com/indices/msci-emerging-markets-historical-data}
        
        \item MOEX index: \url{https://finance.yahoo.com/quote/IMOEX.ME/history/}
        
        \item Spread between two- and ten-year Russian government bonds: 
        \url{https://cbonds.ru/country/Russia_bond/}
    \end{itemize}

\noindent\textbf{South Africa} \citep[see][Table 1]{KHALFAOUI/etal:23}:

    \begin{itemize} 
\item FTSE/JSE Africa All Share index [JALSH]: Bloomberg terminal 

\item MSCI South Africa index [MXSA]: Bloomberg terminal 

\item MSCI South Africa Islamic index [MXSAI]: Bloomberg terminal 

\item Coinbase Ethereum, U.S. Dollars [ETH]: Federal Reserve Bank of St. Louis

\item Coinbase Litecoin, U.S. Dollars [LTC] Federal Reserve Bank of St. Louis 

\item Coinbase Bitcoin, U.S. Dollars [BTC] Federal Reserve Bank of St. Louis 

\item 1-Year Treasury Bill Secondary Market Rate [TB1YR]: Federal Reserve Bank 
of St. Louis 

\item Infectious Disease Equity Market Volatility [FDEMV]: Economic Policy Uncertainty 

\item Twitter Economic Uncertainty [TEU]: Economic Policy Uncertainty 

\item CBOE Volatility Index [VIX]: Bloomberg terminal 

\item CBOE Gold ETF Volatility Index [GVZ]: Bloomberg terminal 

\item Crude Oil ETF Volatility Index [OVX]: Bloomberg terminal  
\end{itemize}

\subsection{Summary statistics}

\begin{table}[H]
    \centering
    \begin{tabular}{c|cccccc} 
    \multicolumn{7}{c}{\textbf{Russia}}\\
         variable& mean& std&  skewness&  kurtosis&  corr$(Y_t; Y_{t-1})$& rank corr\\ \hline 
         RTS& 0.005& 0.071& 1.543&  6.251&  -0.037& -0.046\\ 
         SnP500& 0.001& 0.082& 0.778&  2.902&  0.026& 0.050\\
         USD/CNY& 0& 0.002& 0.547&  4.142&  -0.168& -0.192\\
         2Y bonds& 0.003& 0.031& 4.418&  53.737&  -0.234& 0.229\\
         Sber shares& -0.004& 0.041& -6.305&  63.399&  -0.028& 0.010\\
         Brent& 0.002& 0.026& -0.858& 6.655&  -0.034& -0.056\\
         RUONIA& 0.005& 0.053& 9.208& 117.301&  0.059& 0.032\\
         MSCI EmMar& 0& 0.011& 0.140&  3.627&  0.180& 0.176\\
         MOEX& -0.002& 0.034& -6.759&  88.344&  0.252& 0.010\\
         Spread 2Y-10Y bonds& -0.002& 0.322& 0.667&  15.863&  -0.135& -0.084\\
\hline
        \multicolumn{7}{c}{}\\
        \multicolumn{7}{c}{\textbf{South Africa}}\\
         variable& mean& std&  skewness&  kurtosis&  corr$(Y_t; Y_{t-1})$& rank corr\\ \hline 
         MXSA& 0& 0.015& -0.872&  9.831&  0.097& 0.167\\ 
         JALSH& 0& 0.013& -0.945&  10.972&  0.089& 0.155\\
         BTC& 0.002& 0.043& -1.164&  13.477&  -0.081& -0.043\\
         LTC& 0.001& 0.063& 0.706&  14.844&  -0.054& -0.052\\
         GVZ& 0.002& 0.042& 0.476&  7.363&  -0.007& 0.093\\
         OVX& 0& 0.054& 0.361&  23.076&  0.106& 0.180\\
         \hline 
        \end{tabular}
\end{table}

\end{document}